\documentclass[twocolumn,nofootinbib,10pt]{revtex4-1}

\usepackage{amsmath}
\usepackage{graphicx}
\usepackage[utf8]{inputenc}
\usepackage[usenames, dvipsnames]{color}

\newcommand{\gappeq}{\mathrel{\rlap {\raise.5ex\hbox{$>$}}
{\lower.5ex\hbox{$\sim$}}}}
\newcommand{\lappeq}{\mathrel{\rlap{\raise.5ex\hbox{$<$}}
{\lower.5ex\hbox{$\sim$}}}}

\begin{document}
\title{Dark glueballs and their ultralight axions}
\author{James Halverson} \affiliation{Department of Physics, Northeastern University, Boston, MA 02115-5000 USA} 
\author{Brent D. Nelson} \affiliation{Department of Physics, Northeastern University, Boston, MA 02115-5000 USA} 
\author{Fabian Ruehle} \affiliation{Rudolf Peierls Centre for Theoretical Physics, University of Oxford, Parks Road, Oxford OX1 3PU, UK}
\author{Gustavo Salinas} \affiliation{Department of Physics, Northeastern University, Boston, MA 02115-5000 USA} 

\begin{abstract}
  Dark gauge sectors and axions are
  well-motivated in string theory. We demonstrate that if a confining
  gauge sector gives rise to dark glueballs that are a
  fraction of the dark matter, and the associated axion has a decay constant near the string scale, then this
axion is ultralight and naturally realizes the fuzzy dark matter scenario with
  a modest tuning of a temperature ratio. Astrophysical observations constrain the size of the glueball component relative to the axionic component, while electric dipole moments constrain mixing with the QCD axion.
  
\end{abstract}

\maketitle

Extended gauge and axion sectors are ubiquitous in string theory
and provide natural candidates for hidden sector dark matter. More
specifically, gauge groups
may have high rank, and gauge sectors that interact only via gravity, moduli,
or axions are common. While not all string axions are necessarily associated
with gauge sectors, each gauge sector must couple via its field strength
to some subset of the
axions due to the dynamical nature of the holomorphic gauge coupling function.
It is therefore natural and interesting to consider cosmologies with both hidden
gauge sectors and their associated axions.

We study the possibility that a confining dark gauge sector and its axion both
contribute to the dark matter relic abundance. Previous studies considered either
glueballs or axions that individually satisfied the relic abundance. In the former
case, a single glueball sector may give rise to interesting dark matter
candidates~\cite{Carlson:1992fn,Faraggi:2000pv,Feng:2011ik,Boddy:2014yra,Soni:2016gzf,Acharya:2017szw, Dienes:2016vei, Soni:2016yes, Soni:2017nlm, Draper:2018tmh}, but many glueball sectors (as motivated in string theory)
lead to a dark glueball problem via oversaturation of the relic
abundance~\cite{Halverson:2016nfq}. In the latter case, axions of a variety of
masses and decay constants may constitute the dark matter, and in the specific case in which
$m_a\simeq 10^{-22}$ eV, with decay constant $f_a\simeq 10^{17}$ GeV,
it gives rise to so-called fuzzy dark matter (FDM)~\cite{Hu:2000ke}. FDM has been the subject
of many recent studies~\cite{Hui:2016ltb,Davoudiasl:2017jke,Zhang:2016uiy,Diez-Tejedor:2017ivd,Halverson:2017deq,Murgia:2017lwo,Desjacques:2017fmf}, in part to account for a number of troubling tensions between the standard cold dark matter scenario and observed galactic properties~\cite{Weinberg:2013aya,Tulin:2017ara}.

We demonstrate that if the dark glueball accounts for a sizable fraction
of the dark matter, it is often accompanied by ultralight axionic dark matter which is a candidate for the FDM scenario. This requires a small tuning in the ratio of dark sector to visible sector temperatures, and a solution to the dark glueball problem, but otherwise the scenario arises naturally. Constraints from electric dipole moments may be avoided if the mixing between the dark axion and the QCD axion is minimal.

This paper is organized as follows.
In section~\ref{sec:relics} we compute the relic abundances of dark axions
and glueballs, discuss their dependence on the confinement
scale, and compare these results to the observed dark matter abundance. In section~\ref{sec:param} we discuss the region in parameter space in which the FDM scenario can be realized, while the dark glueball problem is simultaneously eliminated. Constraints on the model arising from astrophysics and electric dipole moment measurements are considered in sections~\ref{sec:astro} and~\ref{sec:EDM}, respectively. Finally, in section~\ref{sec:string}, we discuss the scenario in a string theory context, and consider how many of the assumptions made, and conclusion drawn, are supported by broad arguments in string theory.

\section{Relic densities}
\label{sec:relics}

As in our previous work~\cite{Halverson:2016nfq}, we consider a dark pure Yang-Mills sector (as these are often present in string compactifications) with gauge group $G$ and confinement scale $\Lambda$. In addition, we consider a single axion field $a$ coupling to this sector with the usual $\sim a F \wedge F$ operator, $F$ being the field strength of the hidden gauge field.

We will assume that the dark sector is reheated to a temperature $T_{\text{rh}}' > \Lambda$, such that the dark sector at that time is a thermal bath of dark gluons. Throughout, we will distinguish between the temperature of the observable sector and that of the dark sector by using the symbols $T$ and $T'$, respectively. As the dark sector cools through a transition temperature $T_\Lambda' \sim \Lambda$ the energy density in gluons is converted into glueballs.

These confined states are likely to persist to late times provided that (a) any portals to the Standard Model are generated only by Planck-suppressed effects, and (b) there are no light matter states (analogues to pions in the Standard Model) into which these glueball states can decay. The former is well-motivated by string model-building, while the latter is part of our simplifying assumption. Glueballs can, in principle, decay to cosmological moduli and/or light axions. In~\cite{Halverson:2016nfq}, it was demonstrated that glueball lifetimes will exceed the lifetime of the universe for all realistic UV parameter combinations when $f_a \gtrsim 10^{16}\,{\rm GeV}$. Furthermore, glueball lifetimes from decay into light moduli will be similarly long unless the confinement scale satisfies $\Lambda \gg 1$ GeV. As will be shown in the next section, the confinement scales necessary to generate a FDM axion are of the order $\Lambda \sim 100$ eV. Thus, we should expect dark glueballs to be present cosmologically for the parameter regime of interest. In addition, glueball decays to SM particles are highly suppressed by large powers of $M_{\rm GUT}$, in which case the energy exchange between the visible and dark sectors is minimal and large preferential reheating can be maintained \cite{Forestell:2017wov}.”

After confinement ($T' < T'_\Lambda$), the number density of relic glueballs is set by the rate of $3 \to 2$ interactions, as was studied by Carlson et al.~\cite{Carlson:1992fn}. Freezeout occurs when these interactions cease to be effective, leaving a dark glueball relic, at a decoupling temperature $T_d' \leq\Lambda$. We will assume that throughout these processes, the comoving entropy density is conserved in each sector due to thermal equilibrium and minimal interactions between the sectors, so that the ratio
\begin{equation}
\xi := \frac{s}{s'}
\label{eq:xidef}
\end{equation}
is a constant. We have assumed $T_{\text{rh}}'>\Lambda$, and thus it is reasonable to assume both sectors are relativistic at high enough  $T_{\text{rh}}$. This allows us to identify the parameter $\xi$ as
\begin{equation}
\xi := \frac{g_S T^3}{g_S' T'^3}\, ,
\label{eq:xidef2}
\end{equation}

\noindent with $g_S$ and $g_S^\prime$ being the number of relativistic degrees of freedom in entropy in the visible and dark sectors right after reheating.

As the dark sector is composed of dark gluons at the time of BBN, the number of degrees of freedom in the dark sector has to satisfy the current experimental bounds, given by \cite{Kane:2015qea}
\begin{equation}
g_\star^\prime \vert_{\text{BBN}} \le 148.6 \left(\frac{T_{rh}}{T^\prime_{rh}}\right)^4~.
\label{eq:bbn_constr}
\end{equation}

\noindent This bound can be expressed in terms of the entropy ratio, using (\ref{eq:xidef2}), to give 
\begin{equation}
g^\prime \ge 0.083 \left(\frac{10}{\xi}\right)^4~,
\label{eq:bbn_constr2}
\end{equation}

\noindent if we assume $g_\star^\prime \vert_{\text{BBN}}=g^\prime_{S} \equiv g^\prime$. Notice that mild preferential reheating into the visible sector, $\xi \sim \mathcal{O} (1 - 10)$, is needed to allow for smaller gauge groups.

Since $\xi$ is assumed to be constant, and comparing to the visible sector entropy density today, the relic abundance of glueballs is~\cite{Carlson:1992fn}
\begin{equation}
\Omega_g (\xi) h^2 = \frac{T_d'}{3.6\, {\rm eV}\, \xi}\,.
\label{eq:relicvsTd}
\end{equation}
Following the treatment of~\cite{Halverson:2016nfq}, which updates that of~\cite{Carlson:1992fn}, we can substitute the confinement scale for the decoupling temperature to obtain
\begin{equation}
\Omega_g (\xi) h^2 = \frac{\Lambda}{3.6\, {\rm eV}\, \xi} \omega (\Lambda)\, .
\label{eq:relicvsLambda}
\end{equation}
The multiplicative factor $\omega (\Lambda)$ is simply unity in the case that $3\to 2$ interactions are already decoupled at the time of confinement, where $T_d'\simeq \Lambda$. More typically, we expect $T_d' < \Lambda$, and $3\to 2$ interactions have time to affect the final relic abundance. In this case, assuming radiation domination at the time glueballs are formed\footnote{The case for matter domination is quantitatively similar and has negligible impact on our subsequent results.}, we find
\begin{equation}
\omega (\Lambda) = \frac{4}{5}  \left[W \left(7.45 \times 10^{12}\frac{f^{6/5} g'\,^{4/5}}{\xi^{2/5}} \left(\frac{3.6 {\rm eV}}{\Lambda}\right)^{3/5} \right)\right]^{-1}\, ,
\label{littleomega}
\end{equation}
where $W(x)$ is the Lambert W-function and $f$ is the coupling constant of the effective operator that governs glueball $3 \rightarrow 2$ interactions.

For the axion relic density, we will follow the standard treatment, as contained in~\cite{Sikivie:2006ni,Marsh:2010wq}. We assume the axion mass and scalar potential arise from non-perturbative effects at a time when the universe is radiation dominated. 
The equation of motion for the background field $a(t)$ is given by
\begin{equation}\label{aeqm}
\left(\frac{d^2}{dt^2} + \frac{3}{2t}\frac{d}{dt} \right) a(t) + m_a^2(t) f_a \sin \left(\frac{a(t)}{f_a} \right) = 0\, ,
\end{equation}
where we have used $2t=H^{-1}$ in a radiation-dominated epoch. Let $t_1$ be the time at which the axion field begins to oscillate about its background value. Since the first term scales with $t^{-2}$, this occurs when the second term in~(\ref{aeqm}) grows comparable to the first at a time 
\begin{equation}\notag
m_a(t_1) t_1 \sim 1\, .
\end{equation}
For axion masses of the order $m_a \sim 10^{-22}$ eV, we expect this time to be $t_1 \sim \mathcal{O}(10^{6}\,{\rm sec})$, consistent with the assumption of a radiation-dominated epoch.


%

After the time $t_1$, the axion field evolves adiabatically, with the total number of axions serving as the adiabatic invariant. At this point, the number density for the axion field is
\begin{equation}\notag
n_a = \frac{m_a(t_1) a_i^2}{2} = \frac{a_i^2}{2 t_1}\, ,
\end{equation}
where $a_i$ is the initial amplitude of the field oscillation, and we have taken $m_a(t_1) = t_1^{-1}$ for definiteness. Since the value of $a_i$ is arbitrary, we can parameterize it as $a_i = f_a \theta_i$, where $-\pi \le \theta_i \le +\pi$ is the initial misalignment angle. Therefore, the relic density today would be
\begin{equation}\notag
\rho^0_a = m^0_a n^0_a = m_a^0 \frac{f_a^2 \theta_i^2}{2 t_1} \left(\frac{R_1}{R_0} \right)^3\, ,
\end{equation}
with $R_1$ being the scale factor at $t_1$ and $m^0_a$, $R_0$, the axion mass and scale factor today. Note that we are assuming that the oscillating axion field immediately starts to red-shift as cold dark matter.
If $t_{eq}$ and $R_{eq}$ are the time and scale factor of matter-radiation equality, we have
\begin{align*}
\frac{R_1}{R_{eq}} &= \left(\frac{t_1}{t_{eq}} \right)^{1/2} = \left(\frac{1}{m_a(t_1) t_{eq}} \right)^{1/2}\, ,\\
\frac{R_{eq}}{R_0} &= \left(\frac{t_{eq}}{t_0} \right)^{2/3}\, ,
\end{align*}
which gives
\begin{equation}\notag
\rho^0_a = m_a^0 \frac{f_a^2 \theta_i^2}{2 [m_a(t_1)]^{1/2}} \frac{t_{eq}^{1/2}}{t_0^2}\, .
\end{equation}

Working in Planck units, i.e., $1/m^2_{Pl} = 8 \pi G = 1$, and taking the value of the Hubble constant today to be $H_0 = 1$, we get the relic abundance of axions as
\begin{equation}\notag
\Omega_a = \frac{\rho^0_a}{3} = \frac{m_a^0  f_a^2 \theta_i^2}{6 [m_a(t_1)]^{1/2}} \frac{t_{eq}^{1/2}}{t_0^2}~.
\end{equation}
For an ultralight axion, the mass is assumed to be independent of temperature after oscillations set in, settling at its zero-temperature value long before the axion field starts to oscillate.  Thus we take $m_a(t_1) = m^0_a \equiv m_a$ to obtain
\begin{equation}\notag
\Omega_a = \frac{m_a^{1/2}  f_a^2 \theta_i^2}{6} \frac{t_{eq}^{1/2}}{t_0^2}\, .
\end{equation}
Taking the misalignment angle to be of order one, and using the known values of $t_{eq}$ and $t_0$, we obtain
\begin{equation}\label{Omegaa}
\Omega_a \simeq 0.15 \left(\frac{m_a}{10^{-22}\,{\rm eV}} \right)^{1/2} \left(\frac{f_a}{10^{17}\, {\rm GeV}} \right)^2\, .
\end{equation}
The total relic density in the dark sector is, then,
\begin{equation}\label{Omegatot}
\Omega_{\rm tot}h^2 = \Omega_g (\xi) h^2 + \Omega_a h^2~.
\end{equation}
This is to be compared with the observed relic density~\cite{Aghanim:2015xee} of
\begin{equation}
\Omega_{\rm obs} h^2 = 0.1186 \pm 0.0020\,.
\label{eq:obs_relic}
\end{equation}
Note that only the glueball density has a temperature dependence here.


\section{Dark matter as glueballs and ultralight axions}
\label{sec:param}

The various contributions to~(\ref{Omegatot}) depend on the pair of parameters $\lbrace \Lambda, \xi \rbrace$ for the glueball component, and the pair $\lbrace m_a, f_a \rbrace$ for the axionic component, with a secondary dependence on parameters such as the coupling $f$ and the number of hidden degrees of freedom $g'$. Our ambition is two-fold: (1) to ensure that the total dark matter relic density saturates the value suggested by Planck observations, and (2) to obtain the desirable phenomenology of fuzzy dark matter. 

The latter suggests two further quantitative requirements. First, we wish the total relic density to be dominated by relic axions, not glueballs.\footnote{The glueballs possess self-interactions, therefore realizing the self-interacting dark matter (SIDM) scenario, which is beneficial phenomenologically. However, as will be discussed in section \ref{sec:astro}, the cross sections produced here are much larger than the typical ones for SIDM.} Defining the ratio of densities
\begin{equation}\label{etadef}
\eta \equiv \frac{\Omega_a}{\Omega_g}\, ,
\end{equation}
we desire $\eta$ significantly larger than unity.
Second, we require the axion mass to lie in a narrow range 
\begin{equation}\label{mrange}
10^{-23} \,{\rm eV}  \le m_a \le 10^{-21} \,{\rm eV}\, .
\end{equation}

Achieving these three outcomes simultaneously is not entirely trivial, given that our paradigm assumes a tight relationship between the axion mass and the confinement scale, which in turn sets the mass of the glueball. Specifically, we take the axion mass as generated by Yang-Mills instatons, in which case
\begin{equation} \label{ma}
m_a = \frac{\Lambda^2}{f_a}\, .
\end{equation}
The confinement scale itself, however, is not an independent quantity, given the UV completion of the theory into a string context. Rather, the confinement scale is dynamically generated once the hidden sector gauge group is chosen, and certain boundary conditions are specified. We compute $\Lambda$ via the beta function of super Yang-Mills theory,
which gives
\begin{equation} \label{Lambdaconf}
\Lambda \equiv \Lambda_{\text{IR}} =  \Lambda_{\text{UV}}e^{-2\pi/[3C_2(G)\,\alpha_{\text{UV}}]}\,,
\end{equation}
where $C_2(G)$ is the dual Coxeter number of the gauge group $G$, $\alpha_{\text{UV}}$ is the ultraviolet gauge coupling evaluated at scale $\Lambda_{\text{UV}}$, and $\Lambda_{\text{IR}}$ is the scale at which $\alpha$ diverges. We use the supersymmetric beta functions all the way down to low scale for simplicity.

We thus have a set of three observable quantities, given by~(\ref{Omegatot}), (\ref{etadef})~and~(\ref{ma}), which are governed primarily by three {\it a priori} independent inputs: $\xi$, $f_a$, and $\alpha_{\text{UV}}$. To get a sense of how constrained the problem at hand is, let us take the axion mass and decay constant to be in the intervals (\ref{mrange}) and $10^{15}$ GeV $< f_a < 10^{18}$ GeV. Then, using (\ref{ma}) and (\ref{Lambdaconf}), we obtain the relation
\begin{equation}\label{alphaUV}
\alpha_{\text{UV}} = \frac{4 \pi}{3 C_2(G)} \left[ \ln \left(\frac{\Lambda_{\text{UV}}^2}{m_a f_a}\right)\right]^{-1}.
\end{equation}
For $\Lambda_{\text{UV}}=M_{\text{GUT}}=10^{16}\text{ GeV}$, the mass range in~(\ref{mrange}) corresponds to
\begin{equation}\label{alpharange}
24.2\; C_2(G) \le \alpha_{\text{UV}}^{-1} \le 26.9\; C_2(G)\, .
\end{equation}

The range in~(\ref{alpharange}) is clearly evidence of the exponential sensitivity of the confinement scale, and thus both components of~(\ref{Omegatot}), on the choice of high-scale coupling $\alpha_{\text{UV}}$. Such fine-tuning is a commonly-noted feature of glueball dark matter~\cite{Boddy:2014yra,Soni:2016gzf}. A different choice of $\Lambda_{\text{UV}}$ and/or $C_2$ can alter the range in~(\ref{alpharange}), but the tuning involved is not ameliorated. For comparison with the Standard Model, we note that $\{m_a = 10^{-22}\, {\rm eV}$, $f_a = 10^{17}$ GeV$\}$ gives $\alpha_{\text{UV}} = 1/50.6$, in the case $G=$ SU(2). For the MSSM, with the same $\Lambda_{\text{UV}}$, gauge coupling unification occurs at a value $\alpha_{\text{UV}}^{\rm MSSM} \simeq 1/8\pi \simeq 1/25$, so these values are not unreasonable for a weakly-coupled hidden sector.

\begin{figure}[t]
	\includegraphics[width=.95\columnwidth]{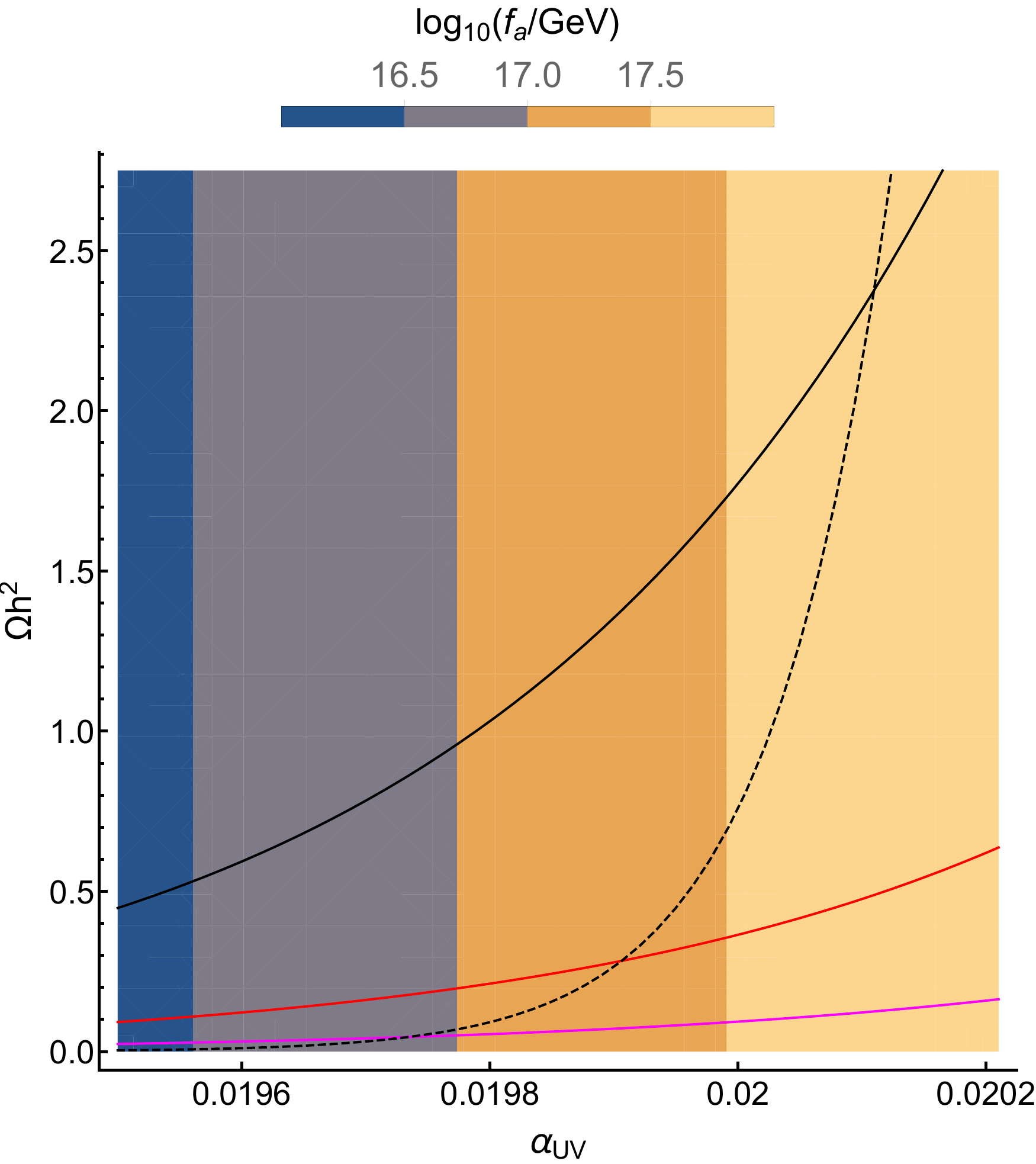}
	\caption{Glueball and axion relic densities for the parameter space of interest. The axion mass is set to $m_a = 10^{-22}$ eV, and the secondary parameters are chosen as $f=0.1$ and $g^\prime=6$ (for SU(2)). The black solid and dashed curves give the glueball (for $\xi=1$) and axion relic density, respectively. The red and magenta curves give the glueball relic density for $\xi = 5$ and $\xi = 20$, respectively.}
	\label{fig:omegaxi}
	\centering
\end{figure}

Restricting our attention to the range of values of $\alpha_{\text{UV}}$ in (\ref{alpharange}) with a hidden sector SU(2) group, we can investigate the relative contributions to $\Omega_{\rm tot}h^2$. Figure \ref{fig:omegaxi} shows the relic densities of glueballs and axions, for the choice $m_a = 10^{-22}$ eV, respectively as the black solid and dashed curves. The glueball relic density far exceeds that of the light axions. What is more, the glueball relic density exceeds the target value in~(\ref{eq:obs_relic}) over the entire range of high-scale couplings, as expected given the dark glueball problem, described at length in~\cite{Halverson:2016nfq}. We note that the parameter $\eta$ in~(\ref{etadef}) varies from $\mathcal{O}(10^{-2})$ to $\mathcal{O}(1)$ over this range in $\alpha_{\text{UV}}$. It is, thus, possible to achieve $\eta>1$ by utilizing the quadratic dependence of $\Omega_a$ on the axion decay constant. This happens when $\alpha_{\text{UV}}$ is larger than its value at the intersection of the glueball and axion curves. Unfortunately, this requires the axion decay constant to be pushed to values uncomfortably close to the Planck scale.

The overall scale of $\Omega_{\rm tot} h^2$ is also too high for both glueball and axionic components. The glueball relic density is mildly sensitive to the size of the self-coupling $f$ in~(\ref{littleomega}); increasing this coupling diminishes the final relic density somewhat, by making depletion mechanisms more effective in the early universe. But this effect is too small to significantly alter our results. So too, increasing the number of degrees of freedom in the hidden sector, $g'$, can diminish the resulting glueball relic density, but achieving the relatively low confinement scale necessary for the fuzzy dark matter outcome typically requires relatively small-rank hidden sector groups, for which we might expect $g'\sim \mathcal{O}(1-10)$.

When confinement scales satisfy $\Lambda \sim \mathcal{O}({\rm 100 ~eV})$, a mild differential in the temperatures of the hidden and observable sectors can reduce the glueball relic density and, potentially, mitigate the dark glueball problem. Significantly, this reduction occurs {\it only} for the glueball relic density -- by our assumptions, the axion relic density is independent of temperature. This effect is demonstrated in Figure~\ref{fig:omegaxi}, where the effect of $\xi$ on the glueball relic density is shown by the red ($\xi = 5$) and magenta ($\xi = 20$) curves. Increasing the value of $\xi$ not only brings down the glueball component, but also decreases the value of $f_a$ for which the axion starts to dominate the relic density, bringing it to one order of magnitude below the Planck scale.

The previous analysis suggests that it should be possible to achieve the three desired observables given in~(\ref{Omegatot}), (\ref{etadef})~and~(\ref{ma}), while keeping the three principal input parameters in theoretically reasonable ranges. Since we are fixing the range of axion masses in~(\ref{mrange}) and $\Lambda \simeq \sqrt{m_a f_a} \sim 100$ eV, the glueball relic density for a given gauge group and boundary condition scale can be seen as a function of $\xi$, $f_a$ and $m_a$
\begin{equation}\notag
\Omega_g \equiv \Omega_g (\xi, f_a, m_a;\, f, g')\, ,
\end{equation}
with a secondary dependence on $f$ and $g'$.

\begin{figure}[t]
\includegraphics[width=.99\columnwidth]{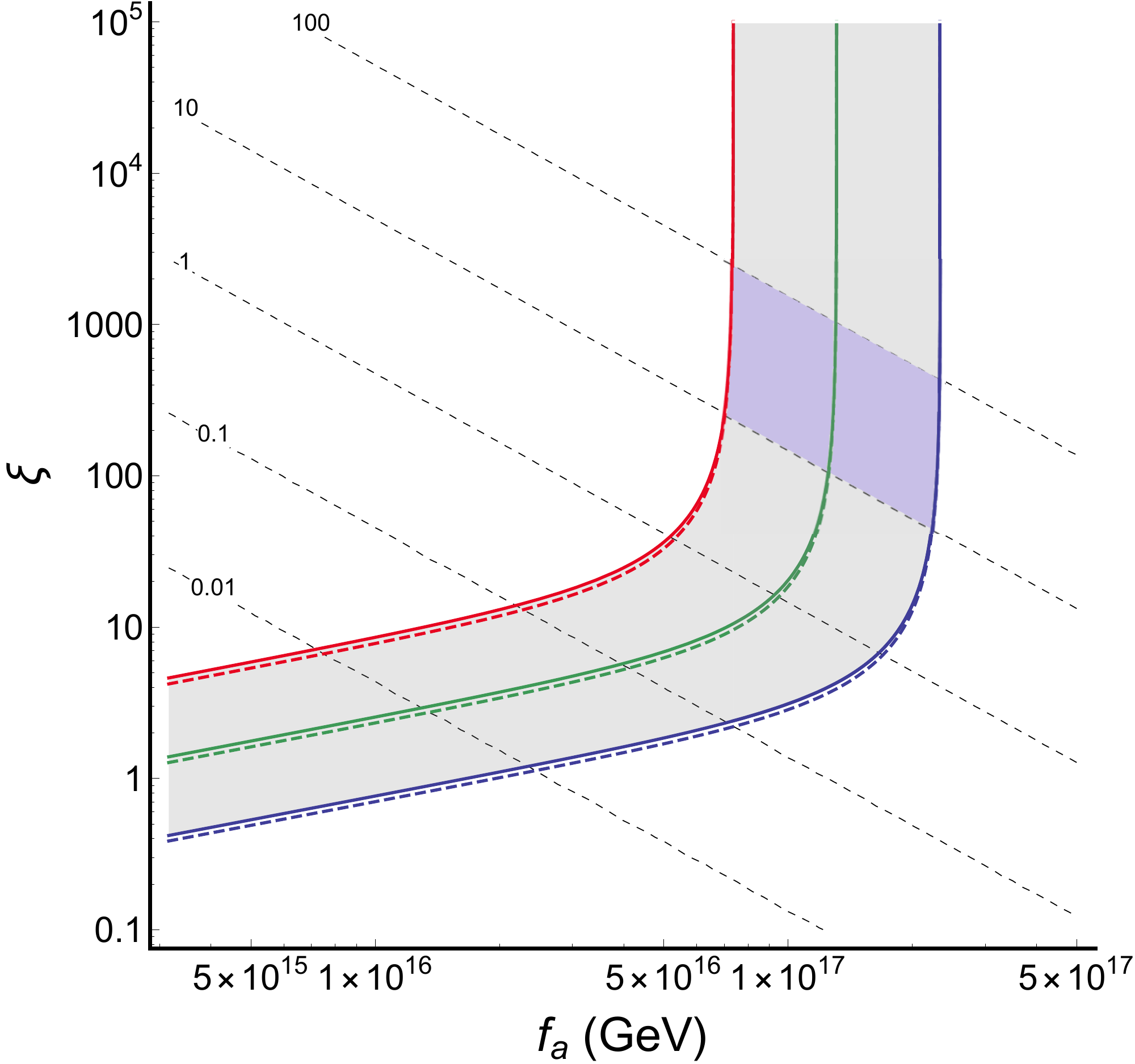}
\caption{Choice of parameters that reproduce the observed DM relic density, as composed of dark glueballs and axions. The green curve represents an axion mass of $m_a = 10^{-22}$ eV, while the blue and red curves represent the boundaries of our range for the axion mass, $m_a = 10^{-23}$ eV and $m_a = 10^{-21}$ eV, respectively. For lower values of $f_a$, the glueball component dominates. Each curve has a maximum allowed value of $f_a$, above which the axion relic density oversaturates (\ref{eq:obs_relic}). Close to this maximum value, the axion component dominates the relic density, as shown by the dashed contours, giving the values of the ratio $\eta$. }
\label{fig:masterplot}
\centering
\end{figure}

This suggests that the locus of points for which $\Omega_{\rm tot} h^2 = \Omega_{\rm obs} h^2 = 0.1186 \pm 0.0020$ with a given axion mass $m_a$ can be plotted in the $\lbrace \xi,\, f_a \rbrace$ plane, as in Figure~\ref{fig:masterplot}. The three solid curves represent the contour where 
$\Omega_{\rm tot} h^2 = \Omega_{\rm obs} h^2$ for axion masses of $m_a = 10^{-21}\,{\rm eV}$ (top, left), $m_a = 10^{-22}\,{\rm eV}$ (central), and $m_a = 10^{-23}\,{\rm eV}$ (bottom, right), respectively. The shaded region in grey thus gives the parameter space in which the total relic density can be made consistent with Planck satellite data, while also generating relic axions that can serve as fuzzy dark matter. 

The third observable is the relative contribution of axions and glueballs to $\Omega_{\rm tot}$. Contours of constant $\eta$ have been superimposed on the shaded area, ranging from $\eta = 0.01$ (bottom left) to $\eta = 100$ (top right). When glueballs dominate ($\eta <1$), the necessary preferential reheating has the dependence $\xi \propto \sqrt{f_a}$, from (\ref{eq:relicvsLambda}) and (\ref{ma}). The curves reach an inflection when $\eta \sim 1$, and become invariant to changes in the temperature ($\xi$) once the axion relic density comes to dominate. In the limit where $\xi,\, \eta \rightarrow \infty$, the allowed range for the axion decay constant is $7.4 \times 10^{16}$ GeV $\le f_a \le 2.3 \times 10^{17}$ GeV. The secondary variables are relevant only in the opposite extreme. For example, the solid curves in Figure~\ref{fig:masterplot} take $g^\prime=1$, whereas the dashed lines take $g^\prime=100$, similar to the number of degrees of freedom of the Standard Model, at high temperature. In the regime where $\eta \gappeq 1$, this mild dependence on internal parameters allows for some flexibility in obtaining the desired phenomenological outcomes. 

Marsh and Silk studied galactic halo models involving a mixture of cold dark matter and axionic dark matter of the variety considered here~\cite{Marsh:2013ywa}. The beneficial effect of fuzzy dark matter for addressing the cusp vs. core problem tended to require that axions make up at least 80-90\% of the halo.\footnote{Given the self-interacting nature of the glueballs, these percentages do not directly translate to our scenario. We, however, expect them to be similar when considering an axion halo with a smaller percentage of glueballs.}  We are thus led to consider the purple region in Figure~\ref{fig:masterplot} for which $10 \le \eta \le 100$. 
Notice that in our preferred region of the $\xi-f_a$ parameter space, one needs a preferential reheating $\xi \sim 10^3$. From the definition (\ref{eq:xidef2}), we obtain
\begin{equation}
	\frac{g_S T_{\rm rh}^3}{g_S' T_{\rm rh}'^3} \sim 10^3 ~\Rightarrow~ \frac{T_{\rm rh}}{T_{\rm rh}^\prime}  \sim 1.6 \, {g^\prime}^{1/3}~,
	\label{Tratio}
\end{equation}
where we have taken $g_S = 228.75$, i.e., we assume that the whole content of the MSSM is relativistic at reheating. Thus, only a factor of $\mathcal{O}(1 - 10)$ is needed for the difference in reheating temperatures to achieve all three desired outcomes simultaneously.

\section{Phenomenological Constraints}
\label{sec:constrain}

\subsection{Astrophysical Observations}
\label{sec:astro}

Fuzzy dark matter is endowed with a fundamental length scale associated with the de~Broglie wavelength of the axion. For masses in the range given by~(\ref{mrange}), this wavelength is on the order of kiloparsecs. Observations of hydrogen absorption lines in the intergalactic medium are sensitive to variations in the matter power spectrum at similar length scales. This suggests that observations of the Lyman-$\alpha$ forest can constrain ultralight axionic dark matter~\cite{Viel:2013apy}.

A number of recent papers have appeared~\cite{Irsic:2017yje,Armengaud:2017nkf,Zhang:2017chj,Kobayashi:2017jcf,Desjacques:2017fmf}, providing conflicting conclusions on precisely where such bounds should be placed. Some authors~\cite{Irsic:2017yje,Armengaud:2017nkf} would interpret the Lyman-$\alpha$ data as forbidding a sizable dark matter component with masses below 1-3 $\times 10^{-21}\,{\rm eV}$, thereby eliminating the entire range in~(\ref{mrange}). However, these interpretations depend on assumptions made when modeling the growth of density perturbations~\cite{Zhang:2017chj,Kobayashi:2017jcf,Desjacques:2017fmf}, and may even depend on details of reionization in the early universe~\cite{DAloisio:2015exk}. We prefer, therefore, to remain agnostic about the precise nature of the Lyman-$\alpha$ bounds, but expect that the red curve in Figure~\ref{fig:masterplot} is likely to be in the least amount of tension with these observations.

For values of $\eta$ in the purple shaded region of Figure~\ref{fig:masterplot}, the fraction of the dark matter relic density accounted for by glueballs is small: a few percent of the total. Nevertheless, this component will be highly self-interacting, and thus astrophysical observations can place meaningful constraints. In the absence of a detailed model of glueball self-interactions, it is customary to employ dimensional analysis, in which one might estimate the glueball interaction cross-section as
\begin{equation} \sigma_{\rm gb} \simeq \frac{4\pi}{\Lambda^2}\, .
\label{xsec}
\end{equation}

Bounds on the self-interactions of dark matter typically come from observations of galaxy mergers, such as the Bullet Cluster~\cite{Randall:2007ph}, or the Abell~520 cluster~\cite{Clowe:2012am}, and generally imply $\sigma_{\text{gb}}/m_{\text{gb}} \lappeq 0.1-1\,{\text{cm}^2/\text{g}}$. In the scenario considered here, the value of the confinement scale is tightly constrained by the desire to obtain a fuzzy dark matter component while simultaneously mitigating the dark glueball problem. As such, a confinement scale $\Lambda = \mathcal{O}(100\,{\rm eV})$ would imply a self-interaction $\sigma_{\text{gb}}/m_{\text{gb}} \gg 1\,{\text{cm}^2/\text{g}}$.
However, extraction of mass-to-light ratios for such cluster mergers comes with some uncertainty, and generally assumes that the interacting dark matter comprises the total of the non-luminous mass. Constraints on a self-interacting sub-component largely disappear once the fraction of the dark matter that self-interacts drops below a few percent~\cite{Boddy:2014yra,Pollack:2014rja}. This is consistent with the values of $\eta$ required
to obtain FDM phenomenology.

\subsection{Electric Dipole Moments}
\label{sec:EDM}

Recently, a search was performed for very light axionic particles, presumed to constitute the bulk of dark matter, by measuring the ratio of spin precession frequencies for ultracold neutrons and $^{199}{\rm Hg}$ atoms~\cite{Abel:2017rtm}. Specifically, the experiment was sensitive to the time-dependent contribution to the combination $d_n - (\mu_n/\mu_{Hg})d_{Hg}$, where the axion-induced neutron electric dipole moment (EDM) is given by
\begin{equation}\notag
d_n(t) \approx 2.4 \times 10^{-16} \frac{\kappa_G a_0}{f_a} \cos(m_a t) ~e\text{-cm}~,
\end{equation}
and the axion-induced $^{199}{\rm Hg}$ EDM is given by
\begin{equation}\notag
d_{Hg}(t) \approx 1.3 \times 10^{-19} \frac{\kappa_G a_0}{f_a} \cos(m_a t) ~e\text{-cm}\, ,
\end{equation}
where $a_0$ is the current amplitude of coherent oscillation of the axion field, $\kappa_G$ is the dimensionless coupling between the ultralight axion and the dual field strength of QCD, and the ratio between the relevant magnetic moments is $\mu_n/\mu_{\rm Hg}= -3.84$.

Assuming that the axions saturate the local energy density of dark matter, such that $\rho \simeq 0.4 ~ \text{GeV}/\text{cm}^3$, with its energy being given by $\rho = m_a^2 a_0^2/2 ~\Rightarrow~ a_0 = \sqrt{2 \rho}/m_a$, we find that the amplitude of the measured combination  $D_0 \equiv \left(d_n - (\mu_n/\mu_{Hg})d_{Hg}\right)_{t=0}$ is given by
\begin{eqnarray}
D_0 & = & 2.4 \times 10^{-16} \frac{\kappa_G a_0}{f_a} ~e\text{-cm} \nonumber \\
 &\simeq& 6 \times 10^3\, \kappa_G \times (10^{-26}~e\text{-cm})\, ,
 \label{D0}
\end{eqnarray}
for the case in which $f_a = 10^{17}\,{\rm GeV}$ and $m_a = 10^{-22}\,{\rm eV}$. 

The experiment looked for an axion-induced oscillation across a large frequency range $10^{-10}\,{\rm Hz} \leq f \leq 10^{-3}\,{\rm Hz}$. The 95\% confidence level upper bound on the amplitude of the oscillation varied somewhat with frequency, but a conservative estimate would be to place and exclusion of $D_0 \leq 
 6 \times 10^{-26}~e$-cm, across the region $10^{-9}\,{\rm Hz} \leq f \leq 10^{-7}\,{\rm Hz}$, which includes the mass range in~(\ref{mrange}).  Thus, the electric dipole moment measurements restrict the value of the coupling to the region $\kappa_G \lesssim 10^{-3}$ for the parameter range of interest here.

We can relax the condition that the ultralight axion saturates the DM relic density and make it contribute a factor $(1+\eta^{-1})^{-1}$ of it, with $\eta$ defined as in the previous section. This would introduce a factor of $(1+\eta^{-1})^{-1/2}$ in (\ref{D0}), which only contributes an $\mathcal{O}(1)$ correction to the limit  $\kappa_G \lesssim 10^{-3}$. If the mixing between the QCD axion and the axion in the dark sector that composes the dark matter is sufficiently small, this constraint for $\kappa_G$ is naturally explained.

\section{String Theory Considerations}
\label{sec:string}

The existence of dark gauge sectors and associated axions with decay constants around $10^{17}$~GeV is well-established in string theory~\cite{Svrcek:2006yi}, and has been most extensively studied in heterotic and F-theory compactifications.

In the heterotic string the 10D gauge group is $E_8\times E_8$ or $SO(32)$ for the supersymmetric string and $SO(16)\times SO(16)$ for the non-supersymmetric string. Consistent compactifications require flux, which results in a breaking of these gauge groups to smaller subgroups. For heterotic orbifolds this was first discussed in~\cite{Ibanez:1986tp,*Ibanez:1987pj} and for free fermionic constructions a discussion can be found in~\cite{Faraggi:1997dc}. For the largest class of models, constructed on smooth Calabi-Yau manifolds with line bundles, millions of models have been found by starting from $E_8\times E_8$~\cite{Anderson:2012yf,Nibbelink:2015vha}, as well as $SO(32)$ and $SO(16)\times SO(16)$ gauge groups~\cite{Nibbelink:2015vha}. By construction, the $E_8\times E_8$ models of~\cite{Anderson:2012yf} have an unbroken hidden sector $E_8$ gauge group, while the models in \cite{Nibbelink:2015vha} always have a hidden sector gauge group of small rank (or a product of small rank gauge groups).

In F-Theory models, there is also significant evidence that large dark gauge sectors are generic~\cite{Grassi:2014zxa,Halverson:2015jua,Taylor:2015ppa,Taylor:2015xtz,Halverson:2016vwx,Halverson:2017ffz,Taylor:2017yqr},
in the sense that non-trivial configurations of intersecting seven-branes that carry non-Abelian gauge groups exist at generic points in complex structure moduli space.

In a concrete string vacuum that realizes a dark glueball and associated 
axion sector, the parameters $(\Lambda,f_a)$ are determined from the ultraviolet
parameters $(\alpha_{\text{UV}},f_a)$ that depend on the VEVs of certain
scalar fields. In type IIB / F-theory compactifications, for example, they depend
on VEVs of complexified K\" ahler moduli. Given an ensemble of metastable
de Sitter vacua across the landscape, one would obtain a distribution of $(\alpha_{\text{UV}},f_a)$
pairs that could either motivate or demotivate the values chosen in this paper.
Concrete large ensembles of metastable de Sitter vacua have not yet been
computed for reasons of complexity, and it is therefore difficult to assess
the likelihood of the $(\Lambda,f_a)$ ranges that we study. However, the possible
existence of these ranges is supported by regime of validity considerations: since
our scenario $\Lambda \ll \Lambda_{QCD}$ 
would arise if the submanifold supporting the dark 
sector is much larger than the one supporting QCD, the supergravity approximation
is likely to be preserved if it already is for the QCD sector. Conversely, if
$\Lambda \gg \Lambda_{QCD}$, which is not the case in our scenario, it would
be more
likely that the supergravity approximation is violated.

As an example, consider the ensemble of $\frac43\times 2.96\times 10^{755}$
F-theory geometries constructed in \cite{Halverson:2017ffz}. Despite the large
number, concrete statements could be made about the geometric gauge group $G$
arising from intersecting seven-branes at generic points in moduli space, due to having
a precise construction algorithm. For example, an exact calculation shows that
$rk(G)\geq 160$ in $\geq 99.9995\%$ of the geometries in the ensemble, and evidence
from random sampling suggests that it is much higher. Note, though, that these gauge groups
all arise from many simple Lie group factors $G_i$, each of which has $rk(G_i)\leq 8$.
It is one of these gauge factors, or alternatively single gauge factors that arise
in other string constructions, that motivate our low energy Yang-Mills sectors.

Much of our analysis has focused on $SU(2)$, and it is natural to ask whether
other common gauge factors in string constructions modify our results. In the context of the mentioned F-theory compactifications, each 
geometric gauge factor $G_i$ is one of the $E_8, E_7, E_6, F_4, SO(8), SO(7), G_2, SU(3),$ or $SU(2)$,
which suggests using $E_8$ and $SU(2)$ as limiting cases. This is well-motivated more broadly
because single gauge factors in string theory often fit inside $E_8$. $G$ affects our analysis
only through $g'=2\,\text{dim}(G)$ and $C_2(G)$, which affects the allowed range of $\alpha_{\text{UV}}$. Since
$C_2(E_8)/C_2(SU(2))=15$ the bounds on $\alpha_{\text{UV}}$  obtain a factor
of $15$ via equation \eqref{alpharange} in passing from $SU(2)$ to $E_8$, and the temperature ratio obtains a factor of $[\text{dim}(E_8)/\rm dim(SU(2))]^{1/3}=[248/3]^{1/3}\simeq 4.5$; both are mild changes.

\section{Conclusion}

We studied a scenario in which the dark matter is composed of dark glueballs and their associated
axions. We found that both components may make up a non-trivial fraction of the observed dark
matter relic abundance if the hidden sector confinement scale is $\Lambda \sim 100$ eV and there is a mild preferential reheating into the visible sector. Remarkably, we find that dark glueballs at this mass scale naturally give rise to ultralight axions and
the fuzzy dark matter scenario, under the assumption that the axion decay constant is
$f_a\simeq O(10^{17} \,\rm GeV)$, as is well-motivated in string theory. The ratio $\eta$ of axion
to glueball relic abundance depends on a few parameters, including the dark confinement scale $\Lambda$,
 the visible to dark sector entropy density ratio $\xi$, and the axion decay constant. We focus primarily
 on $G=SU(2)$, but demonstrate that change of the gauge group between $SU(2)$ and $E_8$ does little to affect the basic results.

Our model is a mixed cosmology of fuzzy dark matter with a self-interacting dark matter component. This presents an opportunity for further study, involving, for example, implications of this scenario on structure formation, as produced by the combination of a component with an intrinsic length scale and another with strong self-interactions. It is also interesting to ask whether or not the Lyman-$\alpha$ forest constraints for FDM are modified in such a scenario.

In addition, we leave a treatment of the effects of axion mixing, which are expected in the context of string theory, to future work. These might have interesting implications, such as constraints on hidden sector physics arising from searches for the QCD~axion, or the generation of new portals between the visible and hidden sectors more generally. The latter might also be useful in addressing the issue of preferential reheating, a common requirement of hidden sector dark matter models such as the ones studied here, as a way of ensuring that our visible sector is always at the highest effective temperature. A systematic study of preferential reheating, considering both details of specific models of reheating, and the subsequent heat exchange between sectors~\cite{Adshead:2016xxj}, would be highly desirable.

\vspace{.2cm}
\noindent \textbf{Acknowledgments.}
J.H.\ is supported by NSF grant PHY-1620526. B.D.N.\ and G.S.\ are supported by NSF grant PHY-1620575. The work of F.R.\ is supported by the EPSRC network grant EP/N007158/1.

\bibliography{temp}

\end{document}